\documentclass[a4paper]{revtex4}
\usepackage{graphicx}
\usepackage{fancyhdr}
\usepackage{amsmath}
\usepackage{color}
\usepackage{graphics}
\usepackage{amsmath, amssymb, epsfig}
\usepackage{subfigure}
\usepackage{graphicx}
\newcommand{\mathsym}[1]{{}}

\newcommand{\ba}{\begin{array}}
\newcommand{\ea}{\end{array}}
\newcommand{\bal}{\begin{align}}
\newcommand{\eal}{\end{align}}
\newcommand{\be}{\begin{equation}}
\newcommand{\ee}{\end{equation}}
\newcommand{\beqa}{\begin{eqnarray}}
\newcommand{\eeqa}{\end{eqnarray}}
\newcommand{ \eq}[1]{Eq.~(\ref{#1})}
\def\321{$SU(3)\times SU(2)\times U(1)$}

\def\ca{\cos\alpha}
\def\cb{\cos\beta}
\def\sa{\sin\alpha}
\def\sb{\sin\beta}

\begin{document}

\title{Limiting two-Higgs-doublet models}

\author{Eung Jin Chun}
\email{ejchun@kias.re.kr }
\affiliation{Korea Institute for Advanced Study, Seoul 130-722, Korea}

\begin{abstract}
Updating various theoretical and experimental constraints on the four different types of two-Higgs-doublet models (2HDMs), we find that only the ``lepton-specific" (or  ``type X") 2HDM can explain the present muon (g-2) anomaly in the parameter region of large $\tan\beta$,
a light CP-odd Higgs boson, and heavier CP-even and charged Higgs bosons which are almost degenerate.
The severe constraints on the models come mainly from the consideration of vacuum stability and perturbativity,  the electroweak precision data, $B$ physics observables like $b\to s \gamma$ as well as the 125 GeV Higgs boson properties measured at the LHC.
\end{abstract}


\maketitle

\thispagestyle{fancy}

\section{Outline}

Since the first measurement of the muon anomalous magnetic moment $a_\mu = (g-2)_\mu/2$
by the E821 experiment at BNL in 2001 \cite{bnl0102}, much progress has been made in both experimental and theoretical sides to reduce the uncertainties by a factor of two or so establishing a consistent $3 \sigma$ discrepancy
\begin{equation}
\Delta a_\mu \equiv a_\mu^{\rm EXP} - a_\mu^{\rm SM} = + 262\, (85) \times 10^{-11}
\end{equation}
which is in a good agreement with the different group's determinations.
Since the 2001 announcement, there have been quite a few studies in the context of 2HDMs  \cite{2hdms,maria0208,cheung0302} restricted only to the type I and II models. 
However, the type X model \cite{cao0909} has some unique features in explaining the $a_\mu$ anomaly while evading all the experimental constraints.

Among many recent experimental results further confirming the Standard Model (SM) predictions,
the discovery of the 125 GeV Brout-Egnlert-Higgs boson, which is very much SM-like,
particularly motivates us to revisit the issue of the muon g$-$2 in favor of the type X 2HDM.

The key features in confronting 2HDMs with the muon g$-$2 anomaly can be summarized as follows \cite{broggio1409,wang1412,abe1504,chun1505}.

\begin{itemize}
\item
The Barr-Zee two loop \cite{bz} can give a dominant (positive) contribution
to the muon g$-$2 for a light CP-odd Higgs boson $A$ and large $\tan\beta$ in the type II and X models.
\item
In the type II model, a light $A$ has a large bottom Yukawa coupling for large $\tan\beta$, and thus is strongly constrained by the collider searches which have not been able to cover a small gap of $~$ 25 (40) GeV $<M_A<$ 70 GeV at the 2 (1) $\sigma$ range of the muon (g-2) explanation \cite{maria0208}.
\item
In the type II (and Y) model, the measured $\bar{B}\to X_s \gamma$ branching ratio pushes
the charged Higgs boson $H^\pm$ high up to 480 (358) GeV at 95 (99) \% C.L. \cite{misiak1503},
which requires a large separation between $M_A$ and $M_{H^\pm}$ putting  a strong limitation on the model due to the $\rho$ parameter bound \cite{cheung0302}.
\item
Consideration of the electroweak precision data (EWPD) combined with the theoretical constraints from the vacuum stability and perturbativity requires the charged Higgs boson almost degenerate with the heavy Higgs boson $H$ \cite{gerard0703} (favoring $M_{H^\pm} > M_H$) and lighter than about 250 GeV in ``the SM limit''; $\cos(\beta-\alpha) \to 0$. This singles out the type X model in favor of the muon g$-$2 \cite{broggio1409}.
\item
In the favored low $m_A$ region, the  125 GeV Higgs decay $h\to AA$ has to suppressed kinematically or by suppressing the trilinear coupling $\lambda_{hAA}$ which is generically order-one.  This excludes the 1 $\sigma$ range of the muon g$-$2 explanation in the SM limit \cite{broggio1409}.
\end{itemize}

However, the latest development \cite{wang1412,abe1504,chun1505} revealed more interesting possibilities in the ``wrong-sign" domain (negative $hbb$ or $h\tau\tau$ coupling) of 2HDMs \cite{ferreira1410}.

\begin{itemize}
\item
A cancellation in $\lambda_{hAA}$ can be arranged to suppress
arbitrarily the $h\to AA$ decay only in the wrong-sign limit with the heavy Higgs masses in the range of $M_{H^\pm} \sim M_H \approx 200 -600$ GeV \cite{wang1412}.
\item
The lepton universality affected by a large $H^+ \tau \nu_\tau$ coupling turns out to severely constrain the large $\tan\beta$ and light $H^\pm$ region of the type X (and II) model and thus  only a very low $M_A$ and $\tan\beta$ region is allowed at 2 $\sigma$ to explain the $a_\mu$ anomaly \cite{abe1504}.
\end{itemize}

\section{Four types of 2HDMs}

Non-observation of flavour changing neutral currents restricts 2HDMs to four different classes which differ by how the Higgs doublets couple to fermions~\cite{Gunion:2002zf}. They are organized by a discrete symmetry $Z_2$ under which different Higgs doublets and fermions carry different parities. These models are labeled as type I, II, ``lepton-specific" (or X) and ``flipped" (or Y).  Having two Higgs doublets $\Phi_{1,2}$, the most general $Z_2$ symmetric scalar potential takes the form:
\begin{eqnarray} \label{scalar-potential}
	V &=& m_{11}^2 |\Phi_1|^2
	+ m_{22}^2 |\Phi_2|^2 - m_{12}^2 (\Phi_1^\dagger \Phi_2 + \Phi_1 \Phi_2^\dagger) \nonumber \\
	&& + {\lambda_1\over2} |\Phi_1|^4 + {\lambda_2 \over 2} |\Phi_2|^4
	+ \lambda_3 |\Phi_1|^2 |\Phi_2|^2 + \lambda_4 |\Phi_1^\dagger \Phi_2|^2 + {\lambda_5 \over 2}
	\left[ (\Phi_1^\dagger \Phi_2)^2 + (\Phi_1 \Phi_2^\dagger)^2\right],
\end{eqnarray}
where a (soft) $Z_2$ breaking term $m^2_{12}$ is introduced.
Minimization of the scalar potential determines the vacuum expectation values $\langle \Phi^0_{1,2} \rangle \equiv v_{1,2}/\sqrt{2}$ around which the Higgs doublet fields 
are expanded as
\begin{equation}
\Phi_{1,2} = \left[\eta^+_{1,2}, {1\over\sqrt{2}}\left(v_{1,2} + \rho_{1,2} + i \eta^0_{1,2}\right)\right].
\end{equation}
The model contains the five physical fields in mass eigenstates  denoted by $H^\pm, A, H$ and $h$.
Assuming negligible CP violation, $H^\pm$ and $A$ are given by
\begin{equation}
 H^\pm, A = s_\beta\, \eta_1^{\pm, 0}  - c_\beta\, \eta_2^{\pm, 0}
\end{equation}
where the angle $\beta$ is determined from $t_\beta\equiv \tan\beta =v_2/v_1$, and their orthogonal combinations are the corresponding Goldstone modes $G^{\pm, 0}$.
The neutral CP-even Higgs bosons are diagonalized as
\begin{equation}
 h = c_\alpha \,\rho_1 - s_\alpha\, \rho_2, \quad
 H = s_\alpha\, \rho_1 + c_\alpha\, \rho_2 
\end{equation}
where $h\, (H)$ denotes the lighter (heavier) state.

The gauge couplings of $h$ and $H$ are given schematically by
$
 {\cal L}_{\rm gauge} = g_V m_V \big(s_{\beta-\alpha} h + c_{\beta-\alpha} H \big) VV
$
where $V=W^\pm$ or $Z$. When $h$ is the 125 GeV Higgs boson, the SM limit corresponds to
$s_{\beta-\alpha} \to 1$.  Indeed, LHC finds,  $c_{\beta-\alpha} \ll 1 $ in all the 2HDMs confirming the SM-like property of the 125 GeV boson \cite{atlas14010}.

\begin{table}[!ht]
\begin{small}
\begin{center}
\begin{tabular}{|l|ccccccccc|}
 \hline
~~~~~~~~&~~~$y_u^A$~~~ & ~~~$y_d^A$~~~ & ~~~$y_l^A$~~~ & ~~~~$y_u^H$~~~ & ~~~$y_d^H$~~~ & ~~~$y_l^H$~~~ & ~~~$y_u^h$~~~ & ~~~$y_d^h$~~~ & ~~~$y_l^h$~~~\\
\hline
~Type I~~ &$\cot\beta$ & $-\cot\beta$ & $-\cot\beta$ & $\frac{\sa}{\sb}$ & $\frac{\sa}{\sb}$ & $\frac{\sa}{\sb}$ & $\frac{\ca}{\sb}$ & $\frac{\ca}{\sb}$ & $\frac{\ca}{\sb}$~ \\
~Type II~~ &$\cot\beta$ & $\tan\beta$ & $\tan\beta$ & $\frac{\sa}{\sb}$ & $\frac{\ca}{\cb}$ & $\frac{\ca}{\cb}$ & $\frac{\ca}{\sb}$ & $-\frac{\sa}{\cb}$ & $-\frac{\sa}{\cb}$~ \\
~Type X~~ &$\cot\beta$ & $-\cot\beta$ & $\tan\beta$ & $\frac{\sa}{\sb}$ & $\frac{\sa}{\sb}$ & $\frac{\ca}{\cb}$ & $\frac{\ca}{\sb}$ & $\frac{\ca}{\sb}$ & $-\frac{\sa}{\cb}$~ \\
~Type Y~~ &$\cot\beta$ & $\tan\beta$ & $-\cot\beta$ & $\frac{\sa}{\sb}$ & $\frac{\ca}{\cb}$ & $\frac{\sa}{\sb}$ & $\frac{\ca}{\sb}$ & $-\frac{\sa}{\cb}$ & $\frac{\ca}{\sb}$~ \\
\hline
\end{tabular}
\end{center}
\end{small}
\caption{The normalized Yukawa couplings for up- and down-type quarks and charged leptons.}
\label{yukawas}
\end{table}

Normalizing the Yukawa couplings of the neutral bosons to a fermion $f$  by $m_f/v$ where $v=\sqrt{v_1^2+v_2^2} = 246$ GeV, we have the following Yukawa terms:
\begin{eqnarray}
-{\cal L}^{\rm 2HDMs}_{\rm Yukawa} &=&
\sum_{f=u,d,l} {m_f\over v}
\left( y_f^h h \bar{f} f + y_f^H H \bar{f} f - i y_f^A A \bar{f} \gamma_5 f \right) \\
&& \nonumber
+\left[ \sqrt{2} V_{ud} H^+  \bar{u} \left( {m_u\over v} y^A_u P_L  + {m_d \over v} y^d_A P_R\right) d
+\sqrt{2} {m_l \over v} y_l^A H^+ \bar{\nu} P_R l + h.c.\right]
\end{eqnarray}
where the normalized Yukawa coupligs $y^{h,H,A}_f$ are summarized in Table I for each of these four types of 2HDMs.

Let us now recall that the tau Yukawa coupling $y_\tau \equiv y^h_l$ in Type X
($y_b \equiv y^h_d$ in Type II)
can be expressed as
\begin{equation} \label{ytau}
y_\tau = -{s_\alpha \over c_\beta} = s_{\beta-\alpha} - t_\beta c_{\beta-\alpha}
\end{equation}
which allows us to have the wrong-sign limit $y_\tau \sim -1$ compatible with the LHC data \cite{ferreira1410}  if $c_{\beta-\alpha}\sim 2/t_\beta$ for large $\tan\beta$ favoured by the muon g$-$2.
Later we will see that a cancellation in $\lambda_{hAA}$ can be arranged only for $y^h_\tau<-1$
to suppress the $h\to AA$ decay.

\section{Electroweak constraints}
\label{sec:ewc}

\begin{figure}[!ht]
\centering
\includegraphics[width=1.0\textwidth]{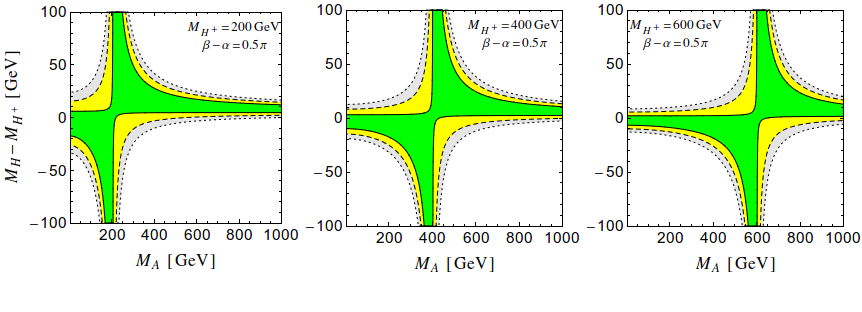}
\caption{The parameter space allowed in the $M_A$ vs.\ $\Delta M_H =M_H-M_{H^\pm}$ plane by EW precision constraints. The green, yellow, gray regions satisfy $\Delta \chi_{\mbox{$\scriptscriptstyle{\rm EW}$}}^2 (M_A, \Delta M)< 2.3, 6.2, 11.8$, corresponding to 68.3, 95.4, and 99.7\% confidence intervals, respectively.}
\label{fig:ewpc}
\end{figure}

Let us fist consider the constraints arising from EWPD on 2HDMs. In particular, we compare the theoretical 2HDMs predictions for $M_W$ and $\sin^2\!\theta^{\text{lept}}_{\text{eff}}$ with their present experimental values via a combined $\chi^2$ analysis. These quantities can be computed perturbatively by means of the following relations
\begin{eqnarray}
	M^2_W &=& \frac{M^2_Z}{2}\left[1 + \sqrt{1 - \frac{4 \pi \alpha_{\rm em}}{\sqrt{2} G_F M^2_Z} \frac{1}{1 - \Delta r}} \, \right]\\
	\sin^2\!\theta^{\text{lept}}_{\text{eff}} &=&
 k_l \! \left(M^2_Z\right) \sin^2\!\theta_W \, ,
\end{eqnarray}
where $\sin^2\!\theta_W=1-M^2_W/M^2_Z$, and $k_l(q^2) = 1+\Delta k_l(q^2)$ is the real part of the vertex form factor $Z\to l \bar{l}$ evaluated at $q^2 = M^2_Z$.
We than use the following experimental values:
\beqa \label{exp}
	M_W^{\mbox{$\scriptscriptstyle{\rm EXP}$}} &=& 80.385 \pm 0.015~{\rm GeV}, \nonumber \\
	\sin^2\!\theta^{\text{lept}, \scriptscriptstyle{\rm EXP}}_{\text{eff}} &=& 0.23153 \pm 0.00016.
	\eeqa
The results of our analysis are displayed in Fig.~\ref{fig:ewpc} confirming a custodial symmetry
limit of our interest $M_A \ll M_H \sim M_{H^\pm}$ (or $M_H \ll M_A\sim M_{H^\pm}$) \cite{gerard0703}.

\section{Theoretical Constraints on the splitting $M_A$-$M_{H^+}$}
\label{sec:thc}

Although any value of $M_A$ is allowed by the EW precision tests in the limit of $M_H \sim M_{H^\pm}$, a large separation between $M_{H^\pm}$ and $M_A$ is strongly constrained by theoretical requirements of vacuum stability, global minimum, and perturbativity:
\begin{eqnarray} \label{vacuum-stability}
&&	\lambda_{1,2}>0,~~\lambda_3>-\sqrt{\lambda_1
	\lambda_2},~~|\lambda_5|<\lambda_3+\lambda_4+\sqrt{\lambda_1 \lambda_2}, \\
&& m_{12}^2(m_{11}^2-m_{22}^2\sqrt{\lambda_1/\lambda_2})(\tan\beta-(\lambda_1/\lambda_2)^{1/4})>0, \\
&&	|\lambda_i| \lesssim |\lambda_{\rm max}| =  \sqrt{4\pi}, 2\pi,\mbox{ or } 4\pi .
\end{eqnarray}
Taking $\lambda_1$ as a free parameter, one can have the following expressions for the other couplings in the large $t_\beta$ limit \cite{chun1505}:
\begin{eqnarray} \label{lambdas}
\lambda_2 v^2 &\approx& s^2_{\beta-\alpha} M_h^2  \\
\lambda_3 v^2 &\approx& 2 M^2_{H^\pm}
-(s^2_{\beta-\alpha} + s_{\beta-\alpha} y_\tau) M_H^2 + s_{\beta-\alpha} y_\tau M_h^2\\
\lambda_4 v^2 &\approx&  -2 M^2_{H^\pm} + s^2_{\beta-\alpha} M^2_H + M_A^2 \\
\lambda_5 v^2 &\approx&  s^2_{\beta-\alpha} M^2_H - M_A^2 
\end{eqnarray}
where we have used the relation (\ref{ytau}) neglecting the terms of
${\cal O}(1/t_\beta^2)$.

%
\begin{figure}[!ht]
\centering
\subfigure{\includegraphics[width=0.3\textwidth]{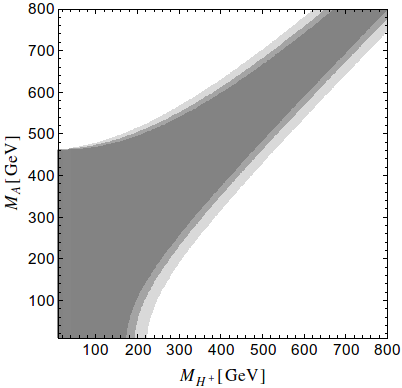}}\quad
\subfigure{\includegraphics[width=0.3\textwidth]{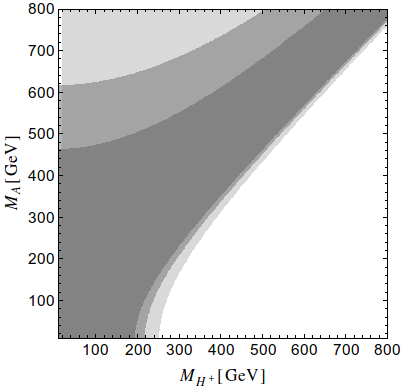}}
\caption{Theoretical constraints on the $M_A$-$M_{H^\pm}$ plane.
  The darker to lighter gray regions in the left panel correspond to the allowed regions for $\Delta M\equiv M_H-M_{H^\pm}= \{20,0,-30\}$ GeV and $\lambda_{\rm max} = \sqrt{4 \pi}$. The allowed regions in the right panel correspond to $\lambda_{\rm max} = \{\sqrt{4 \pi}, 2 \pi, 4 \pi\}$ and vanishing $\Delta M$. 
  }
\label{fig:semiana}
\end{figure}

Consideration of  all the theoretical constraints mentioned above in the SM limit
corresponding to $s_{\beta-\alpha}=y_\tau =1$ gives us Fig.~\ref{fig:semiana}.
One can see that for a light pseudoscalar with $M_A \lesssim 100$~GeV the charged Higgs boson mass gets an upper bound of $M_{H^\pm} \lesssim 250$~GeV.

\section{Constraints from the muon g$-$2}
\label{sec:gm2}
Considering all the updated SM calculations of the muon g$-$2,  we obtain
\be
  a_{\mu}^{\mbox{$\scriptscriptstyle{\rm SM}$}}= 116591829 \, (57) \times 10^{-11}
\ee
comparing it with the experimental value
$   a_{\mu}^{\mbox{$\scriptscriptstyle{\rm EXP}$}}  = 116592091 \, (63) \times 10^{-11}$,
one finds a deviation at 3.1$\sigma$:
$\Delta a_{\mu} \equiv a_{\mu}^{\mbox{$\scriptscriptstyle{\rm EXP}$}}-a_{\mu}^{\mbox{$\scriptscriptstyle{\rm SM}$}} = +262 \, (85) \times 10^{-11}$.
In the 2HDM, the one-loop contributions to $a_{\mu}$ of the neutral and charged Higgs
bosons are
\be
	\delta a_\mu^{\mbox{$\scriptscriptstyle{\rm 2HDM}$}}({\rm 1loop}) =
	\frac{G_F \, m_{\mu}^2}{4 \pi^2 \sqrt{2}} \, \sum_j  \left (y_{\mu}^j \right)^2  r_{\mu}^j \, f_j(r_{\mu}^j),
\label{amuoneloop}
\end{equation}
where $j =  \{h, H, A , H^\pm\}$, $r_{\mu}^ j =  m_\mu^2/M_j^2$, and
\begin{eqnarray}
	f_{h,H}(r) &=& \int_0^1 \! dx \,  { x^2 ( 2- x) \over 1 - x + r x^2},
\label{oneloopintegrals1} \\
	f_A (r) &=& \int_0^1 \! dx \,  { -x^3  \over 1 - x +r  x^2},
\label{oneloopintegrals2} \\
	f_{H^\pm} (r) &=& \int_0^1 \! dx \, {-x (1-x)  \over 1 - (1-x) r}.
\label{oneloopintegrals3}
\end{eqnarray}
These formula show that the one-loop contributions to $a_{\mu}$ are positive for the neutral scalars $h$ and $H$, and negative for the pseudo-scalar and charged Higgs bosons $A$ and $H^{\pm}$ (for $M_{H^\pm} > m_{\mu}$). In the limit $r\ll1$,
\begin{eqnarray}
	f_{h,H}(r) &=&- \ln r - 7/6 + O(r),
	\label{oneloopintegralsapprox1} \\
	f_A (r) &=& +\ln r +11/6 + O(r),
	\label{oneloopintegralsapprox2} \\
	f_{H^\pm} (r) &=& -1/6 + O(r),
	\label{oneloopintegralsapprox3}
\end{eqnarray}
showing that in this limit $f_{H^\pm}(r)$ is suppressed with respect to $f_{{h,H,A}}(r)$.
Now the two-loop Barr-Zee type diagrams with effective
$h\gamma \gamma$, $H\gamma \gamma$ or  $A\gamma \gamma$ vertices generated
by the exchange of heavy fermions gives
\be
	\delta a_\mu^{\mbox{$\scriptscriptstyle{\rm 2HDM}$}}({\rm 2loop-BZ}) = \frac{G_F \, m_{\mu}^2}{4 \pi^2 \sqrt{2}} \, \frac{\alpha_{\rm em}}{\pi}
	\, \sum_{i,f}  N^c_f  \, Q_f^2  \,  y_{\mu}^i  \, y_{f}^i \,  r_{f}^i \,  g_i(r_{f}^i),
\label{barr-zee}
\end{equation}
where $i = \{h, H, A\}$, $r_{f}^i =  m_f^2/M_i^2$, and $m_f$, $Q_f$ and $N^c_f$ are the mass, electric charge and number of color degrees of freedom of the fermion $f$ in the loop. The functions $g_i(r)$ are
\be
\label{2loop-integrals}
	g_i(r) = \int_0^1 \! dx \, \frac{{\cal N}_i(x)}{x(1-x)-r} \ln \frac{x(1-x)}{r},
\ee
where ${\cal N}_{h,H}(x)= 2x (1-x)-1$ and ${\cal N}_{A}(x)=1$.

\begin{figure}[!ht]
\centering
\subfigure{\includegraphics[width=0.3\textwidth]{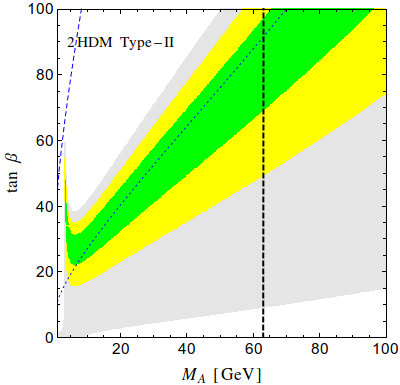}}\quad
\subfigure{\includegraphics[width=0.3\textwidth]{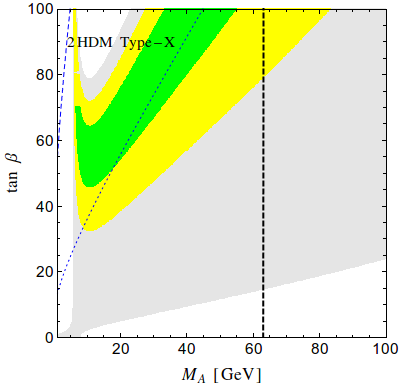}}
\caption{The $1\sigma$, $2\sigma$ and $3\sigma$ regions allowed by $\Delta a_\mu$ in the
$M_A$-$\tan\beta$ plane taking the limit of $\beta-\alpha=\pi/2$ and $M_{h(H)}=126$ (200) GeV in
type II (left panel) and type X (right panel) 2HDMs. The regions below the dashed (dotted) lines are allowed at 3$\sigma$ (1.4$\sigma$) by $\Delta a_e$. The vertical
dashed line corresponds to $M_A = M_h/2$. }
\label{fig:g-2}
\end{figure}

Note the enhancement factor $m_f^2/m_{\mu}^2$ of the two-loop formula in~\eq{barr-zee} relative to the one-loop contribution in~\eq{amuoneloop},
which can overcome the additional loop suppression factor $\alpha / \pi$, and makes
the two-loop contributions may  become larger than the one-loop ones.
Moreover, the signs of the two-loop functions $g_{h,H}$ (negative) and
$g_{A}$ (positive) for the CP-even and CP-odd contributions are
opposite to those of the functions $f_{h,H}$ (positive) and $f_{A}$ (negative) at one-loop.
As a result, for small $M_A$ and large $\tan \beta$ in Type II and X, the positive two-loop pseudoscalar contribution can generate a dominant contribution which can account for
the observed $\Delta a_{\mu}$ discrepancy.
The additional 2HDM contribution $\delta a_{\mu}^{\mbox{$\scriptscriptstyle{\rm 2HDM}$}}  = \delta a_\mu^{\mbox{$\scriptscriptstyle{\rm 2HDM}$}}({\rm 1loop}) + \delta a_\mu^{\mbox{$\scriptscriptstyle{\rm 2HDM}$}}({\rm 2loop-BZ})$ obtained adding Eqs.~(\ref{amuoneloop}) and~(\ref{barr-zee}) (without the $h$ contributions) is compared with $\Delta a_\mu$ in Fig.~\ref{fig:g-2}.
%

Finally, let us remark that the $hAA$ coupling is generically order one and thus can leads to a sizable non-standard decay of $h \to A A$ which should be suppressed kinematically or by making $|\lambda_{hAA}/v|\ll1$ to meet the LHC results \cite{chun1505,wang1412,abe1504}.
Using Eq.~(\ref{lambdas}), one gets the $hAA$ coupling, $\lambda_{hAA}/v \approx s_{\beta-\alpha} [\lambda_3+\lambda_4-\lambda_5]$, and thus
\begin{equation}
\lambda_{hAA} v/s_{\beta-\alpha} \approx
-(1+s_{\beta-\alpha} y_\tau) M_H^2 + s_{\beta-\alpha} y_\tau M_h^2 + 2 M_A^2
\end{equation}
where we have put $s^2_{\beta-\alpha}=1$ \cite{chun1505}. It shows that,  in the SM limit of $s_{\beta-\alpha} y_\tau \to 1$, the condition $\lambda_{hAA} \approx 0$ requires $M_H \sim M_h$ which is disfavoured, and thus one needs to have $M_A > M_h/2$. On the other hand, one can arrange a cancellation for $\lambda_{hAA} \approx 0$ in the wrong-sign domain $s_{\beta-\alpha} y_\tau <0$ if the tau Yukawa coupling satisfies
\begin{equation}
y_\tau s_{\beta-\alpha} \approx - {M_H^2 - 2 M_A^2 \over M_H^2-M_h^2}.
\end{equation}

\section{Summary}
The type X 2HDM provides a unique opportunity to explain the current $\sim 3 \sigma$ deviation in the muon g$-$2  while satisfying all the theoretical requirements and the experimental constraints. The parameter space favourable for the muon g$-$2 at 2$\sigma$ is quite limited in the SM limit: $\tan\beta \gtrsim 30$ and $M_A \ll M_H \sim M_{H^\pm} \lesssim 250$ GeV.
However, consideration of the $h\to AA$ decay and lepton universality \cite{abe1504} rules out this region. On the other hand, in the wrong-sign limit of $y_\tau \sim -1$, a cancellation for $\lambda_{hAA}\approx 0$ can be arranged for $M_H$ up to about 600 GeV \cite{wang1412,chun1505} opening up more parameter space.

Such a light CP-odd boson $A$ and the extra Heavy bosons can be searched for at the next run of the LHC mainly through $pp \to  A H, AH^\pm$ followed by the decays $H^\pm \to \tau^\pm \nu$ and $A,H\to \tau^+ \tau^-$ \cite{abe1504,chun1505}.


\end{document}